\documentclass[a4paper,fleqn,usenatbib]{mnras}

\usepackage[T1]{fontenc}
\usepackage{ae,aecompl}

\usepackage{graphicx}	
\usepackage{amsmath}	
\usepackage{amssymb}	
\usepackage{multirow}
\usepackage{tabularx}
\usepackage{threeparttable}

\title[Maximum CR energy in hotspots is not determined by 
synchrotron cooling]{Evidence that the maximum
electron energy in hotspots of FR~II galaxies is not determined by 
synchrotron cooling}

\author[A. T. Araudo et al.]{
Anabella T. Araudo,$^{1}$\thanks{E-mail: Anabella.Araudo@physics.ox.ac.uk}
Anthony R. Bell,$^{2}$
Aidan Crilly$^{1,3}$
and Katherine M. Blundell$^{1}$
\\
$^{1}$University of Oxford, Astrophysics, Keble Road, 
Oxford OX1 3RH, UK\\
$^{2}$University of Oxford, Clarendon Laboratory, Parks Road, 
Oxford OX1 3PU, UK\\
$^{3}$University of Cambridge, Downing College, Regent Street, 
Cambridge CB2 1DQ, UK
}

\date{Accepted XXX. Received YYY; in original form ZZZ}

\pubyear{2015}

\begin{document}
\label{firstpage}
\pagerange{\pageref{firstpage}--\pageref{lastpage}}
\maketitle

\begin{abstract}
It has been suggested that relativistic  shocks in extragalactic
sources may accelerate the highest energy cosmic rays. The maximum energy 
to which cosmic rays can be accelerated  depends on the structure of 
magnetic turbulence near the shock but recent 
theoretical advances indicate that relativistic shocks are probably 
unable to 
accelerate particles to energies much larger than a PeV.
We study the hotspots of powerful radiogalaxies, where
electrons accelerated at the termination shock emit synchrotron radiation.
The turnover of the synchrotron spectrum is typically observed between 
infrared and optical frequencies,
indicating that the maximum energy of non-thermal electrons accelerated
at the shock is $\lesssim$~TeV for a canonical magnetic field of
$\sim$100~$\mu$G. 
Based on theoretical considerations we show that this maximum energy cannot be 
constrained by synchrotron losses as usually assumed, unless the jet density
is unreasonably large and most of the jet upstream energy goes to 
non-thermal particles. We test this result by considering a sample of hotspots 
observed with high spatial resolution at radio, infrared and optical 
wavelengths.
\end{abstract}



\section{Introduction}

Active Galactic Nuclei (AGN) have been proposed as sources of 
Ultra High Energy Cosmic Rays (UHECRs). Shocks with different velocities and
extents are present in jets of Fanaroff-Riley (FR) radiogalaxies
\citep{FR}, where 
particles can be accelerated via diffusive shock acceleration.
In particular, relativistic and mildly relativistic shocks with
velocity $v_{\rm sh}$ at the jet termination region 
might accelerate particles with  Larmor radius
$r_{\rm g} \sim R_{\rm j}$, where $R_{\rm j} \sim 1$~kpc is the jet width.
Particles with such a large $r_{\rm g}$ in a magnetic field
$\sim 100$~$\mu$G have energy 
\begin{equation}
\frac{E_{\rm UHECR}}{\rm EeV} \sim 100
\left(\frac{v_{\rm sh}}{c/3}\right)
\left(\frac{B}{100 \,\rm \mu G}\right)
\left(\frac{R_{\rm j}}{\rm kpc}\right),
\label{hillas}
\end{equation}
as expected for UHECRs \citep{Lagage-Cesarsky,Hillas}. In particular,
\cite{Rachen_93} and \cite{Norman_95} concluded that hotspots of FR~II 
radiogalaxies are plausible sources of UHECRs 
\cite[see also][]{Nagano_Watson_00,Kotera_Rv_11}.
But, there are two assumptions behind Eq.~(\ref{hillas}): 1) particles
diffuse in the Bohm regime, i.e. the mean-free path is $\lambda \sim r_{\rm g}$,
and 2) the magnetic field $B$ persists over distances $\sim R_{\rm j}$
downstream of the shock. 

Protons are the dominant component of
UHECRs. Given that ion radiation losses are slow in low density
plasmas such as AGN jets, protons can be accelerated up to energies 
$E_{p,\rm max} \sim E_{\rm UHECR}$ if both assumptions are satisfied. 
However, there are no hadronic radiative signatures from hotspots and 
therefore we do not have any observational information about $E_{p,\rm max}$.
In consequence, we investigate the validity of  assumptions 1) and 2) 
by modelling the synchrotron emission produced by
non-thermal electrons accelerated at the jet reverse shock.
The synchrotron turnover at $\nu_{\rm c} \gtrsim 10^{14}$~Hz typically observed 
in hotspots of FR~II galaxies 
\cite[e.g.][]{3c273-Natur,Meisenheimer_IR,Tavecchio_05,Stawarz_07,Werner_12}
indicates that the maximum energy of non-thermal 
electrons is 
\begin{equation}
\frac{E_{\rm c}}{\rm TeV} \sim 0.2
\left(\frac{\nu_{\rm c}}{10^{14}\,{\rm Hz}}\right)^{\frac{1}{2}}
\left(\frac{B}{100\,\mu{\rm G}}\right)^{-\frac{1}{2}}
\label{Ec}
\end{equation}
\citep{Ginzburg}, much smaller than $E_{\rm UHECR}$ for reasonable values 
of the magnetic field. 
The traditional assumption is that $E_{\rm c}$ is determined by synchrotron 
cooling and therefore the diffusion coefficient of particles with 
such an energy is 
\begin{equation}
\frac{\mathcal{D_{\rm c,s}}}{\mathcal{D_{\rm Bohm}}} \sim 10^7
\left(\frac{v_{\rm sh}}{c/3}\right)^2
\left(\frac{\nu_{\rm c}}{10^{14}\,{\rm Hz}}\right)^{-1}
\label{DDBohm}
\end{equation}
\cite[e.g.][]{Stage_06,Kirk_Brian_10}. Protons with  energy $\sim$$E_{\rm c}$ 
also diffuse with $\mathcal{D_{\rm c,s}}$
and therefore the maximum energy that they can 
achieve is reduced to 10~TeV instead of 100~EeV as expected from the Hillas
constraint in Eq.~(\ref{hillas}), assuming that $B$ persists over distances
larger than the synchrotron cooling length $l_{\rm c}$ of  electrons with energy 
$E_{\rm c}$. However, theoretical results of Weibel-mediated
shocks \cite[e.g.][]{Spitkovsky_08I}, and observational
analysis of the case study of 4C74.26 \citep{4c7426} indicate that 
the magnetic field is damped in the downstream region of a relativistic  shock.

Numerical simulations show that Weibel-mediated  shocks in relativistic 
and weakly magnetised plasmas amplify the magnetic field on 
scale length $s \ll r_{\rm g}$, where $r_{\rm g}$ is the Larmor radius of 
particles being accelerated \citep{Spitkovsky_08I,Spitkovsky_08II}. 
The mean-free path for
scattering by small-scale turbulence ($\lambda \sim r_{\rm g}^2/s$) is larger 
than the size of the system ($R_{\rm j}$ in our case) if $s$ is comparable 
with the plasma skin depth  $c/\omega_{\rm pi}$ and $r_{\rm g}$
is the Larmor radius of an EeV proton.  
These small-scale magnetic fluctuations decay
at a distance $\sim$100~$c/\omega_{\rm pi}$ downstream of the shock,
corresponding to $\sim$10$^{11}$~cm for a mildly relativistic plasma
with density of the order of $10^{-4}$~cm$^{-3}$ (see Eq.~(\ref{c_omega_pi})).
This rapid decay of the fluctuations inhibits particle acceleration 
to EeV energies, as was pointed out by  
\cite{lemoine-pelletier-10, Sironi_13} and \cite{Brian_14}.

Damping of the magnetic field in the downstream region of a relativistic 
shock was observationally confirmed by modelling the jet termination
region of the quasar 4C74.26 \citep{4c7426}. 
The compact synchrotron emission ($\sim$0.1~kpc)  detected in the 
southern hotspot of this source would require 
a magnetic field $\sim 2.4$~mG to match the size of the emitter with 
the synchrotron cooling length at the observed frequency of 1.66~GHz. 
This value of the magnetic field is about 10 times the upper limit imposed by
the equipartition condition with non-thermal particles (see more details
of the model in \cite{heproV}). Therefore, the compact radio
emission delineates the region within which the magnetic
field is amplified by plasma instabilities up to
$\sim$100~$\mu$G, and it is damped
downstream of the shock. 

In \cite{4c7426} we consider the standard framework to explain the
cut-off of the synchrotron spectrum at IR/optical 
frequencies and we discussed the 
thickness of the synchrotron emitter in the context of Weibel instabilities.
Given that the thickness of the MERLIN radio emitter in the southern 
hotspot of 4C74.26 is larger than the 
synchrotron cooling length of $E_{\rm c}$-electrons, we interpreted this 
behaviour of $E_{\rm c}$ being determined by synchrotron cooling, and then, at
distances $\sim$0.1~kpc downstream of the shock, the magnetic field is damped
as a consequence of the small scale of Weibel turbulence.
However, $\sim$0.1~kpc is
much larger than the turbulence decay length predicted by numerical
simulations of Weibel-mediated shocks in plasmas with densities 
$\sim$10$^{-4}$~cm$^{-3}$.

In the present work we carry out a deeper study of particle acceleration
in the hotspots of FR~II radiogalaxies. We
revisit the assumption of the synchrotron turnover being
determined by synchrotron losses. Given that the scale-length of 
magnetic fluctuations
has to be larger than $c/\omega_{\rm pi}$ (see Sect.~\ref{mfa}),  we show that
$E_{\rm c}$ cannot be determined by synchrotron cooling, as usually assumed, 
unless the jet density is unreasonably large and most of the jet upstream
energy goes to non-thermal particles.
We also show that the Weibel 
instability is not the source of the amplified magnetic field throughout the
whole hotspot emission region since not only does it damp too quickly, 
but also it generates turbulence on a very small scale,
insufficient to accelerate particles up to $E_{\rm c}\sim$TeV for typical 
values of the magnetic field. 
In Sect.~\ref{hotspots_mfd} we discuss the alternative possibility that
the magnetised turbulence is generated by the Non Resonant Hybrid 
instability \citep{Bell_04} which damps less quickly and grows on a 
larger scale.

In Table~\ref{argument} we present some of the mathematical relations we use
and how the reigning paradigm violates energy conservation even with very 
conservative assumptions. We consider the sample of hotspots observed with 
high spatial resolution at radio, infrared (IR) and optical frequencies 
in \cite{Mack_09}. We find that very low values of the magnetic field,
and therefore a huge energy density in non-thermal electrons, would be 
required to explain the flux density at 8.4~GHz if
the IR/optical cut-off of the synchrotron spectrum was constrained by 
synchrotron cooling (see Sect.~\ref{syn_constraint}). 
These results invite the revision of previous phenomenological models of the
hotspots non-thermal emission.

The results presented in this paper 
have also important implications for Eq.~(\ref{hillas}) and
the maximum energy that protons can achieve by being accelerated in 
the jet reverse shock. 
We conclude that  hotspots of FR~II radiogalaxies with optical 
synchrotron cut-off are very poor accelerators of UHECRs.

The paper is organised as follows: In Sect.~\ref{hotspots} we introduce 
the reader to state-of-the-art modelling of non-thermal particles
in hotspots.
In Sect.~\ref{mfa} we revise the assumption that the acceleration process
ceases as a consequence of efficient synchrotron losses and show that 
this standard picture is in disagreement with a limit imposed by plasma 
physics. In Sect.~\ref{plasma} we show that $\mathcal{D}_{\rm c,s}/\mathcal{D}_{\rm Bohm}\sim 10^6-10^7$
cannot be explained in the framework of any known instability.
In Sect.~\ref{hotspots_mfd} we explore a possible scenario to constrain the 
maximum energy of particles accelerated in the jet reverse shock,
and in Sect.~\ref{disc} we present our conclusions. Throughout the 
paper we use cgs units and the cosmology 
$H_0 = 71$~km~s$^{-1}$~Mpc$^{-1}$, $\Omega_0 = 1$ and $\Lambda_0 = 0.73$.

\begin{table*}
\caption{Key features of the argument showing that extreme densities 
in the jet plasma would be required if the cut-off of the synchrotron spectra
were determined by synchrotron cooling, and our new conjecture for an
alternative explanation.}
\begin{threeparttable}
\label{argument}
\begin{tabularx}{1\textwidth}{c|c|X}
\hline
\multirow{7}{*}{Plasma physics} 
&\multirow{2}{*}{$r_{\rm g} = \frac{E}{eB}$} &Larmor radius of particles with 
energy $E$ and  electric charge $e$ in a magnetic field $B$\\
&\multirow{2}{*}{$\mathcal{D} = \lambda \frac{c}{3},\,\,\,\lambda =  \frac{r_{\rm g}^2}{s}$} & 
Diffusion coefficient $\mathcal{D}$ and mean-free path $\lambda$ of particles 
in a medium  with magnetic turbulence of scale length $s$\\
&\multirow{1}{*}{$c/\omega_{\rm pi} = c/\sqrt{4\pi n_{\rm j} e^2/\bar\gamma_p m_p}$}
&Ion skin depth in a jet with density $n_{\rm j}$ and  
mean energy $\bar\gamma_pm_pc^2$\\
&\multirow{1}{*}{$s \ge \frac{c}{\omega_{\rm pi}}$ $\Rightarrow \lambda \le \lambda_{\rm max} \equiv \frac{r_{\rm g}^2}{c/\omega_{\rm pi}}$} & $c/\omega_{\rm pi}$ is the smallest 
characteristic plasma scale-length\tnote{a} 
(see Sect.~\ref{mfa} and Eq.~(\ref{rg_ratio}))\\
\hline
\multirow{3}{*}{Diffusive shock acceleration} 
&$t_{\rm acc} \sim 20 \mathcal{D}/v_{\rm sh}^2$& 
Acceleration timescale in a parallel shock with velocity $v_{\rm sh}$\\
&\multirow{2}{*}{$U_e<U_{\rm kin}$}&  The energy density in 
non-thermal electrons (see Eq.~(\ref{Ue}))
cannot be greater than the energy budget $U_{\rm kin}$ (see Sect.~\ref{hotspots} 
and Eq.~(\ref{U_kin})) \\
\hline
\multirow{3}{*}{Synchrotron radiation} 
&\multirow{2}{*}{$\gamma = 4.5\times10^{-4}\sqrt{\nu/B}$}& {Lorentz factor of 
electrons emitting synchrotron photons with frequency $\nu$}\\
&$t_{\rm synchr} = 7.5\times10^8/(\gamma B^2)$& 
Synchrotron cooling time\\
\hline
\multirow{2}{*}{Observations} & 
$\nu_{\rm c} =10^{14}-10^{15}$~Hz & Cut-off of the 
synchrotron spectrum \\
&$E_{\rm c}=\gamma_{\rm c}m_ec^2=\gamma(\nu_{\rm c})m_ec^2$ & Non-thermal electrons'
maximum energy (see Eqs.~(\ref{Ec}) and (\ref{gamma_c}))\\
\hline
\multirow{2}{*}{Reigning paradigm} 
&$t_{\rm acc}(\gamma_{\rm c}) = t_{\rm synchr}(\gamma_{\rm c})$&
Synchrotron losses govern where the cut-off is \\
&$\lambda_{\rm c,s} \propto v_{\rm sh}^2\nu_{\rm c}^{-1/2} B^{-3/2}$  
& Mean-free path of 
$\gamma_{\rm c}$-electrons (see Sect.~\ref{model_to_date} and Eq.~(\ref{mfp_c}))\\
\hline	
\multirow{2}{*}{Combining the above}  & 
\multirow{2}{*}{$n_{\rm j}>10^{-5}$-$10^{-4}$~cm$^{-3}$}
& A very large jet density 
is required to be
$\lambda_{\rm c,s}\le\lambda_{\rm max}$ and $U_{\rm e}<U_{\rm kin}$ 
(see Sect.~\ref{syn_constraint} and Table~\ref{tab_sources})
\\
\hline\hline
\multirow{4}{*}{Our conjecture}  & \multirow{2}{*}{
$\lambda(E_{\rm c},B) \le r_{\rm g}(E_{\rm c},B_{\rm jd})$}
& Condition for particle acceleration in a perpendicular shock with magnetic
field $B_{\rm jd}$\\
&\multirow{2}{*}{
$E_{\rm nrh} = E_{\rm c}\frac{B_{\rm jd}}{B}$}&
Maximum energy at which non-thermal protons excite non-resonant turbulence 
(see Sect.~\ref{hotspots_mfd})\\
\hline 
\end{tabularx}
\begin{tablenotes}
\small
\item[a] In electron-positron plasmas, $s$ has to be greater
than the electron-skin depth $c/\omega_{\rm pe}$, where 
$c/\omega_{\rm pe} = \sqrt{m_e/m_p}\,c/\omega_{\rm pi} \sim 0.02\,c/\omega_{\rm pi}$. 
\end{tablenotes}
\end{threeparttable}
\end{table*}

\section{Hotspots}
\label{hotspots}

\begin{figure}
\includegraphics[width=0.8\columnwidth]{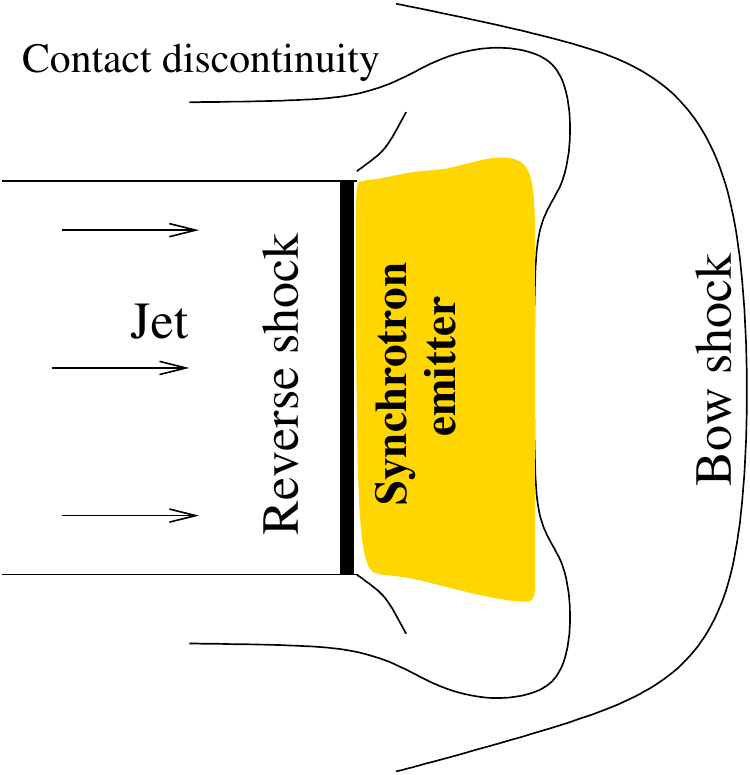}
\caption{Sketch of the standard picture of the jet termination region. 
Particles are accelerated
at the reverse shock, and radiate in the shock downstream region, here
labelled ``Synchrotron emitter''.}
\label{sketch}
\end{figure}

The jet termination region in FR~II radiogalaxies is characterised by a 
double shock structure separated by a contact discontinuity, as sketched in 
Figure~\ref{sketch}. Note however that the contact discontinuity is
unstable due to the velocity shear and density contrast in both
sides of the discontinuity \cite[e.g.][]{Mizuta_04}.
Hotspots are the  downstream region of the 
jet reverse shock, where particles accelerated by the shock emit
synchrotron radiation.

\subsection{Energy budget}
\label{energy_budget}

The  kinetic energy density of relativistic jets with particles of mass $m$ 
and density $n_{\rm j}$ and moving with bulk Lorentz factor $\Gamma_{\rm j}$ is 
\begin{equation}
\frac{U_{\rm kin}}{\rm erg~cm^{-3}} = 9\times10^{-9} 
\left(\frac{\Gamma_{\rm j} - 1}{0.06}\right)
\left(\frac{n_{\rm j}}{10^{-4}\,{\rm cm^{-3}}}\right)
\left(\frac{m}{m_p}\right), 
\label{U_kin}
\end{equation}
where $\Gamma_{\rm j} = 1.06$ corresponds to a jet velocity $v_{\rm j} = c/3$
\citep{Casse_05,Steenbrugge_08} and $m_p$ is the proton mass. 
Even in the case that we do not know the jet matter composition,
we expect that ions (from the jet formation 
region  or from entrainment as the jet propagates) 
dominate the jet dynamics at the termination region and therefore $m=m_p$ in
Eq.~(\ref{U_kin}). The jet magnetisation parameter is defined as
\begin{equation}
\sigma_{\rm j} \equiv \frac{U_{\rm mag,j}}{U_{\rm kin}} 
\sim 4.4\times10^{-6}\left(\frac{B_{\rm j}}{\mu\rm G}\right)^2
\left(\frac{\Gamma_{\rm j} - 1}{0.06}\right)^{-1} 
\left(\frac{n_{\rm j}}{10^{-4}\,{\rm cm^{-3}}}\right)^{-1},
\label{sigma}
\end{equation}
where $U_{\rm mag,j}=B_{\rm j}^2/8\pi$ and $B_{\rm j}$ is the jet's magnetic field.
The jet (upstream) ram pressure is converted into thermal, 
non-thermal and magnetic
($U_{\rm mag} = 4\times10^{-10}(B/100\,\mu{\rm G})^2$~erg~cm$^{-3}$) pressure
in the shock downstream region with magnetic field $B$.
The magnetic field in the jet downstream region cannot be greater than
\begin{equation}
\frac{B_{\rm sat}}{100 \,\rm \mu G} = 4.8
\left(\frac{\Gamma_{\rm j} - 1}{0.06}\right)^{\frac{1}{2}} 
\left(\frac{n_{\rm j}}{10^{-4}\,{\rm cm^{-3}}}\right)^{\frac{1}{2}},
\label{l_cool}
\end{equation}
\citep{3c273-Natur} which corresponds to the extreme case  
$U_{\rm mag} = U_{\rm kin}$.
The jet density is unknown in most cases, 
but $2\times10^{-4}$~cm$^{-3}$ is the upper-limit for the primary hotspot in the 
Western lobe of
Cygnus~A given the non-detection of radio polarisation \citep{polarization},
and $6\times10^{-5}$~cm$^{-3}$ is the upper limit in 3C273 \citep{3c273-Natur}.

\subsection{Model to date}
\label{model_to_date}

Hotspot (radio-to-optical) synchrotron spectra typically show a cut-off
at $\nu_{\rm c} \gtrsim 10^{14}$~Hz
\cite[e.g.][]{Meisenheimer_IR,Tavecchio_05,Zhang_10,Werner_12}. 
The traditional assumption is that 
the maximum energy of non-thermal electrons accelerated at the jet reverse 
shock, $E_{\rm c} = \gamma_{\rm c}m_ec^2$, where
\begin{equation}
\gamma_{\rm c} \sim 4.5\times10^5 
\left(\frac{\nu_{\rm c}}{10^{14}\,{\rm Hz}}\right)^{\frac{1}{2}}
\left(\frac{B}{100\,\mu{\rm G}}\right)^{-\frac{1}{2}},
\label{gamma_c}
\end{equation}
is determined by a competition between synchrotron cooling 
and acceleration timescales (see Table~\ref{argument}).
By equating $t_{\rm synchr}(\gamma_{\rm c}) = t_{\rm acc}(\gamma_{\rm c})$,
the diffusion coefficient $\mathcal{D_{\rm c,s}}$ is given by Eq.~(\ref{DDBohm})
and the mean-free path of the $\gamma_{\rm c}$-electrons is
\begin{equation}
\frac{\lambda_{\rm c,s}}{\rm pc} \sim 25
\left(\frac{v_{\rm sh}}{c/3}\right)^{2}
\left(\frac{\nu_{\rm c}}{10^{14}\,{\rm Hz}}\right)^{-\frac{1}{2}}
\left(\frac{B}{100\,\mu{\rm G}}\right)^{-\frac{3}{2}}.
\label{mfp_c}
\end{equation}
These electrons  radiate half of their energy over a distance 
\begin{equation}
\frac{l_{\rm c}}{\rm kpc} 
\sim 0.02 \left(\frac{r}{7}\right)^{-1}
\left(\frac{\nu_{\rm c}}{10^{14}\,{\rm Hz}}\right)^{-\frac{1}{2}}
\left(\frac{B}{100\,{\rm \mu G}}\right)^{-\frac{1}{5}}
\left(\frac{v_{\rm sh}}{c/3}\right)
\label{l_c}
\end{equation}
downstream of the shock, where we have assumed that the velocity of the
shocked plasma is $v_{\rm d} = v_{\rm sh}/r$, being $4\lesssim r \lesssim 7$
the adiabatic shock compression factor. Note however that our results are not 
sensitive to the exact value of $r$. The condition 
$t_{\rm synchr}(\gamma_{\rm c}) = t_{\rm acc}(\gamma_{\rm c})$ implies that
the size of the acceleration region is
$L_{\rm acc} \sim l_{\rm c}$. 

\begin{figure}
\includegraphics[width=0.52\textwidth]{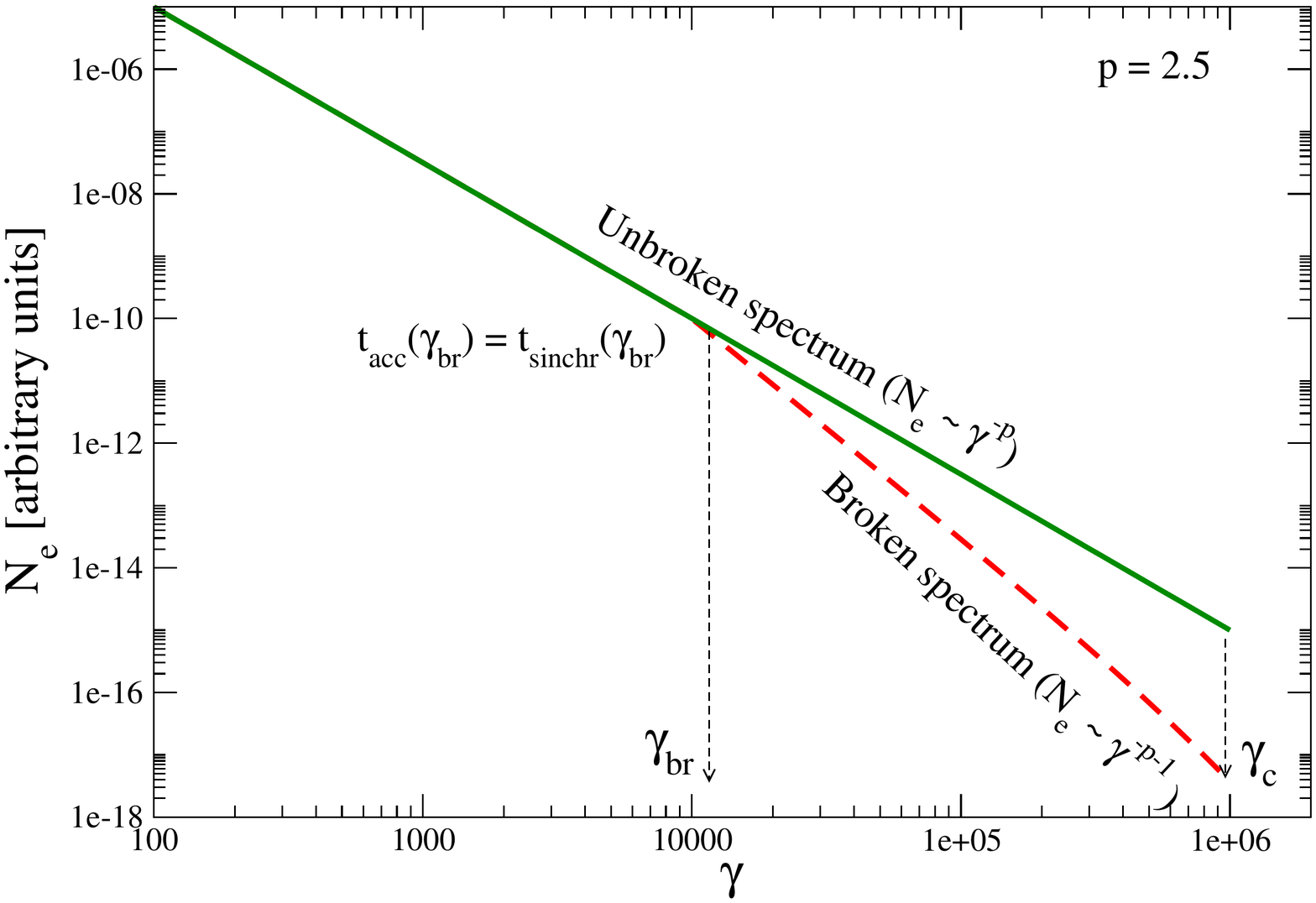}
\includegraphics[width=0.52\textwidth]{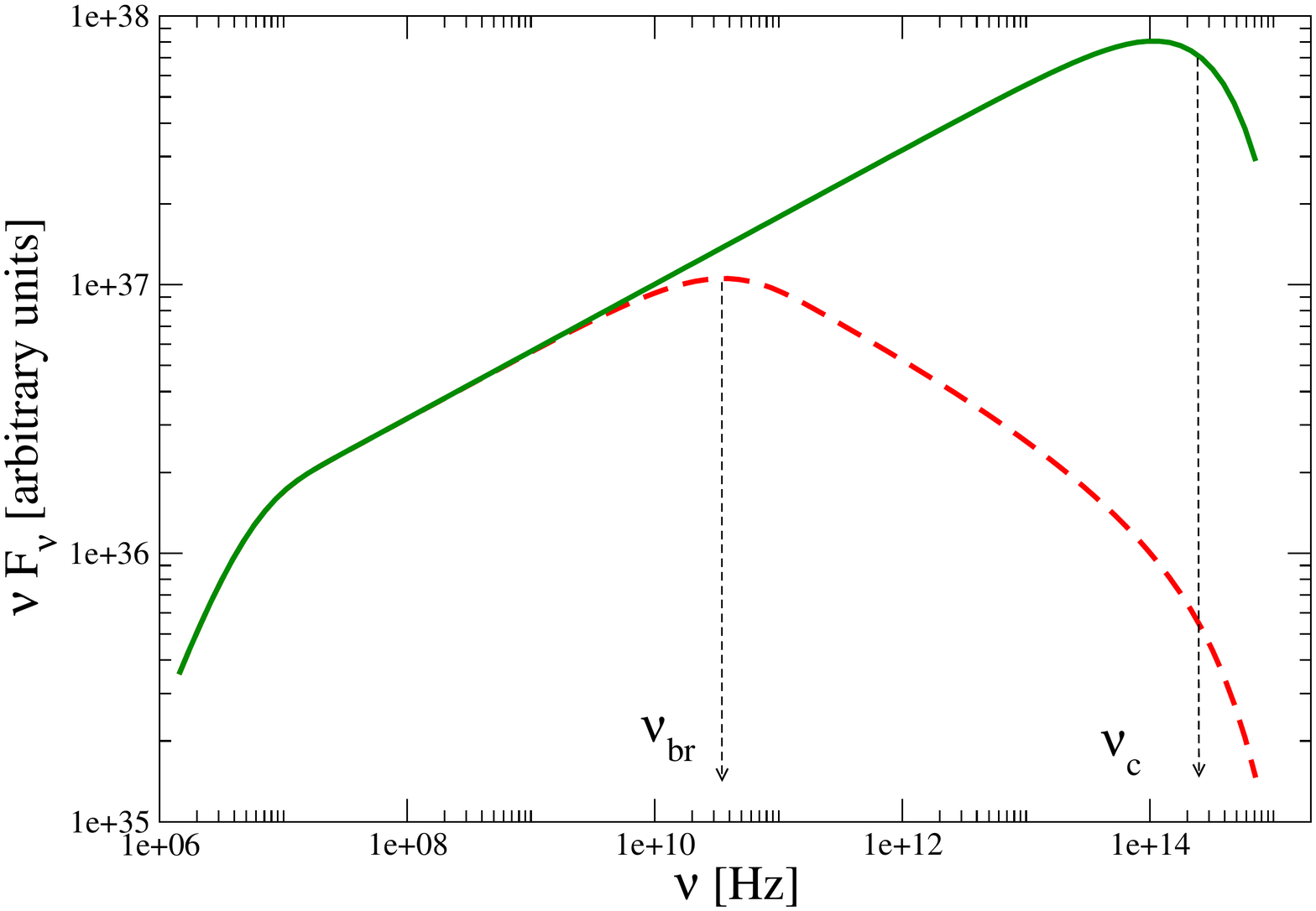}
\caption{\emph{Top:} Non-thermal electron energy distributions 
for the case where accelerated particles are injected in the shock
downstream region following a power-law distribution $\propto \gamma^{-p}$,
where $p=2.5$ (see Sect.~\ref{syn_constraint}).
\emph{Bottom:} Synchrotron spectra. The break $\gamma_{\rm br}$ and cut-off 
$\gamma_{\rm c}$ in $N_e$ correspond to $\nu_{\rm br}$ and $\nu_{\rm c}$ in the
synchrotron spectrum.}
\label{spectra}
\end{figure}

In some cases, the spectrum is broken at frequency 
$\nu_{\rm br}$.  
To avoid misunderstandings between $\nu_{\rm br}$
and $\nu_{\rm c}$ we show in Fig.~\ref{spectra} two canonical electron and 
synchrotron spectra: broken (red-dashed lines) and unbroken 
(green-solid lines). 
\cite[See e.g.][for a comparison with real spectra.]{3c273-Natur}
In sources with enough radio-to-optical data to be able to fit the synchrotron
spectrum and measure $\nu_{\rm br}$ and $\nu_{\rm c}$, the magnetic 
field is determined by comparing the synchrotron cooling time  at 
$\nu_{\rm br}$ with the timescale $L/v_{\rm d}$ to be the particles advected 
a distance $L$ from the shock \cite[e.g.][]{3c273-Natur}:
\begin{equation}
\frac{B}{\mu{\rm G}} \sim 354\left(\frac{r}{7}\right)^{-\frac{2}{3}}
\left(\frac{\nu_{\rm br}}{10\,{\rm GHz}}\right)^{-\frac{1}{3}}
\left(\frac{v_{\rm sh}}{c/3}\right)^{\frac{2}{3}}
\left(\frac{L}{{\rm kpc}}\right)^{-\frac{2}{3}}.
\label{B_breack}
\end{equation}
Therefore, by replacing $B$ in Eq.~(\ref{mfp_c}), the mean free path of 
the most energetic electrons accelerated at the shock is
\begin{equation}
\frac{\lambda_{\rm c,s}}{L} 
\sim 0.05 \left(\frac{r}{7}\right)^{-1}
\left(\frac{\nu_{\rm br}}{\nu_{\rm c}}\right)^{\frac{1}{2}}
\left(\frac{v_{\rm sh}}{c/3}\right).
\label{lambda_s0}
\end{equation}
The main uncertainty is $L$, that depends on the angle $\theta_{\rm j}$
between the jet and the line of sight through the equation
\begin{equation}\label{L_theta}
L  = \frac{l_{\rm br} - D\cos\theta_{\rm j}}{\sin{\theta_{\rm j}}},
\end{equation}
where $l_{\rm br}$ is the observed size at $\nu_{\rm br}$ and $D\sim 2R_{\rm j}$ 
is the diameter
of the source (when hotspots are modelled as cylinders of thickness $L$). 
Note that when the jet lies on the plane of the sky, $\theta_{\rm j} = 90^{\circ}$
and $L = l_{\rm br}$. 

In the seminal paper of \cite{Meisenheimer_89}, using observations 
at optical, near IR, millimetre and radio bands, hotspots are classified into 
\emph{high loss} ($\nu_{\rm br} \le 10$~GHz) and \emph{low loss} 
($\nu_{\rm br} \gg 10$~GHz) sources. The latter are characterised
by thin emission regions with
$L\sim$0.13 ($D/L = 5.85$), 0.06 ($D/L = 22$), and 0.07~kpc ($D/L = 28.4$)
in the sources 3C20~West, 3C33~South and 3C111~East, respectively. 
In these \emph{low loss} sources $B \sim 0.1 B_{\rm eq}$\footnote{Magnetic 
fields below the equipartition value $B_{\rm eq}$ are also found in 
hotspots where the X-ray emission is also modelled
\cite[e.g.][]{Zhang_10,Werner_12}. \cite{Werner_12} mentioned that this 
behaviour is in agreement with \cite{DeYoung_02}, who showed that magnetic 
field amplification by magnetohydrodynamic turbulence to equipartition 
values requires timescales greater than the dwell
time of the plasma in the hotspots, unless special conditions are
imposed.}  
and $\lambda_{\rm c,s}\sim$1-8~pc, where $B_{\rm eq}$ is the magnetic field 
in equipartition with non-thermal particles. 
The thin (disc-like) emission regions in 
these hotspots were suggested to be the result of a drastic change in 
the downstream flow, producing a rapid decay of the magnetic field.

The detection of diffuse IR and optical synchrotron emission on scales 
larger than $l_{\rm c}$
has been interpreted as in-situ re-acceleration (Fermi~II)
of the non-thermal electrons 
\citep{Meisenheimer_IR,3C445_02Sci,Brunetti_03}. Later on, 
\cite{Mack_09} presented high spatial resolution observations at near-IR,
optical and radio frequencies of low-power radio hotspots, finding that 
$\nu_{\rm br} \sim \nu_{\rm c} \sim 10^{14}-10^{15}$~Hz in all of them
(i.e. \emph{low loss} sources). The cooling time of electrons emitting
synchrotron radiation at these high frequencies in a magnetic field
$\sim 10-100$~$\mu$G (in equipartition with non-thermal electrons and 
protons)
is $\sim 2-5\times10^3$~yr and much shorter than the 
timescales of adiabatic expansion (see Table~5 in \cite{Mack_09}).

In the next section we demonstrate
that the maximum energy at which electrons are accelerated cannot
be constrained by synchrotron losses, as usually assumed. 
To demonstrate this, we consider a sample of hotspots 
that do not show a break in their synchrotron  spectra
(green-solid lines in Fig.~\ref{spectra}), but our arguments are not 
restricted to these sources.

\section{Revising the  synchrotron cut-off: when observational 
astronomy meets plasma physics}
\label{mfa}

The synchrotron turnover at $\nu_{\rm c} \gtrsim 10^{14}$~Hz  observed in
hotspots of FR~II radiogalaxies indicates that the maximum energy of 
non-thermal electrons accelerated at the jet reverse shock is 
$E_{\rm c} = \gamma_{\rm c} m_ec^2$, where $E_{\rm c}$ and $\gamma_{\rm c}$ are
given by Eqs.~(\ref{Ec}) and (\ref{gamma_c}), respectively.
The Larmor radius of these particles is 
\begin{equation}
\frac{r_{\rm g}(\gamma_{\rm c})}{\rm cm} \sim 9\times10^{12} 
\left(\frac{\nu_{\rm c}}{10^{14}\,{\rm Hz}}\right)^{0.5}
\left(\frac{B}{100\,\mu{\rm G}}\right)^{-1.5}
\end{equation}
and the mean-free path is $\lambda \sim s/\theta^2
\sim r_{\rm g}^2/s$, where $\theta \sim s/r_{\rm g}$ is the 
deflection angle of particles interacting with magnetic inhomogeneities of
scale length  $s$. 
Considering the jet as a hydrogen plasma with electron and proton 
thermal Lorentz factors
$\bar\gamma_{e} $ and $\bar\gamma_{p}\sim \Gamma_{\rm j}$, 
respectively, the ion skin depth downstream of the shock is
\begin{equation}
\frac{c}{\omega_{\rm pi}} \sim 8.6\times10^8 \sqrt{\Gamma_{\rm j}}
\left[\left(\frac{r}{7}\right)\left(\frac{n_{\rm j}}{10^{-4}\,{\rm cm^{-3}}}\right)\right]^{-\frac{1}{2}}\,\,{\rm cm}.
\label{c_omega_pi}
\end{equation}
The ratio
\begin{equation}
\frac {r_{\rm g}(\bar\gamma_{e})}{c/\omega_{\rm pi}}= \left(\frac{\bar\gamma_e m_e c^2}{\Gamma_{\rm j} m_p c^2} 
\right) \sigma_{\rm j}^{-\frac{1}{2}} \sim 2\frac{\bar\gamma_{e}}{\Gamma_{\rm j}}
\left[\left(\frac{r}{7}\right)^3\left(\frac{\sigma_{\rm j}}{10^{-6}}\right)\right]^{-\frac{1}{2}}
\label{rg_ratio}
\end{equation}
shows that the thermal electron Larmor radius is generally larger than 
$c/\omega_{\rm pi}$ (in the ``hot electrons/cold protons'' scenario)
in which case
$c/\omega_{\rm pi}$ is the smallest characteristic plasma scalelength.
Therefore, considering that $s \ge c/\omega_{\rm pi}$ for suprathermal 
particles, we find an upper-limit $\lambda_{\rm max}$ to the mean-free path of
the most energetic electrons accelerated at the jet reverse shock:
\begin{eqnarray}\label{lambda_c}
\begin{aligned}
\lambda_{\rm max} = 
\frac{r_{\rm g}^2(\gamma_{\rm c})}{c/\omega_{\rm pi}} 
&\sim 0.02
\left(\frac{\nu_{\rm c}}{10^{14}\,{\rm Hz}}\right)
\left(\frac{B}{100\,\mu{\rm G}}\right)^{-3}\\
&\left[\left(\frac{r}{7}\right)\left(\frac{n_{\rm j}}{10^{-4}\,{\rm cm^{-3}}}\right)\right]^{\frac{1}{2}}\,{\rm pc},
\end{aligned}
\end{eqnarray}
independent of the shock velocity $v_{\rm sh}$ (see Table~\ref{argument}). 
Therefore, the maximum diffusion  coefficient is given by
\begin{eqnarray}
\begin{aligned}
\frac{\mathcal{D}_{\rm max}}{\mathcal{D}_{\rm Bohm}} = 
\frac{\lambda_{\rm max}}{r_{\rm g}(\gamma_{\rm c})}
=& 3.2\times10^4
\left(\frac{\nu_{\rm c}}{10^{14}\,{\rm Hz}}\right)^{\frac{1}{2}}
\left(\frac{B}{100\,\mu{\rm G}}\right)^{-\frac{3}{2}}\\
&\left[\left(\frac{r}{7}\right)
\left(\frac{n_{\rm j}}{10^{-4}\,{\rm cm^{-3}}}\right)\right]^{\frac{1}{2}}.
\label{D_DBohm}
\end{aligned}
\end{eqnarray}

\subsection{Is the maximum energy of non-thermal electrons constrained by 
synchrotron losses?}
\label{syn_constraint}

\begin{figure}
\includegraphics[width=0.56\textwidth]{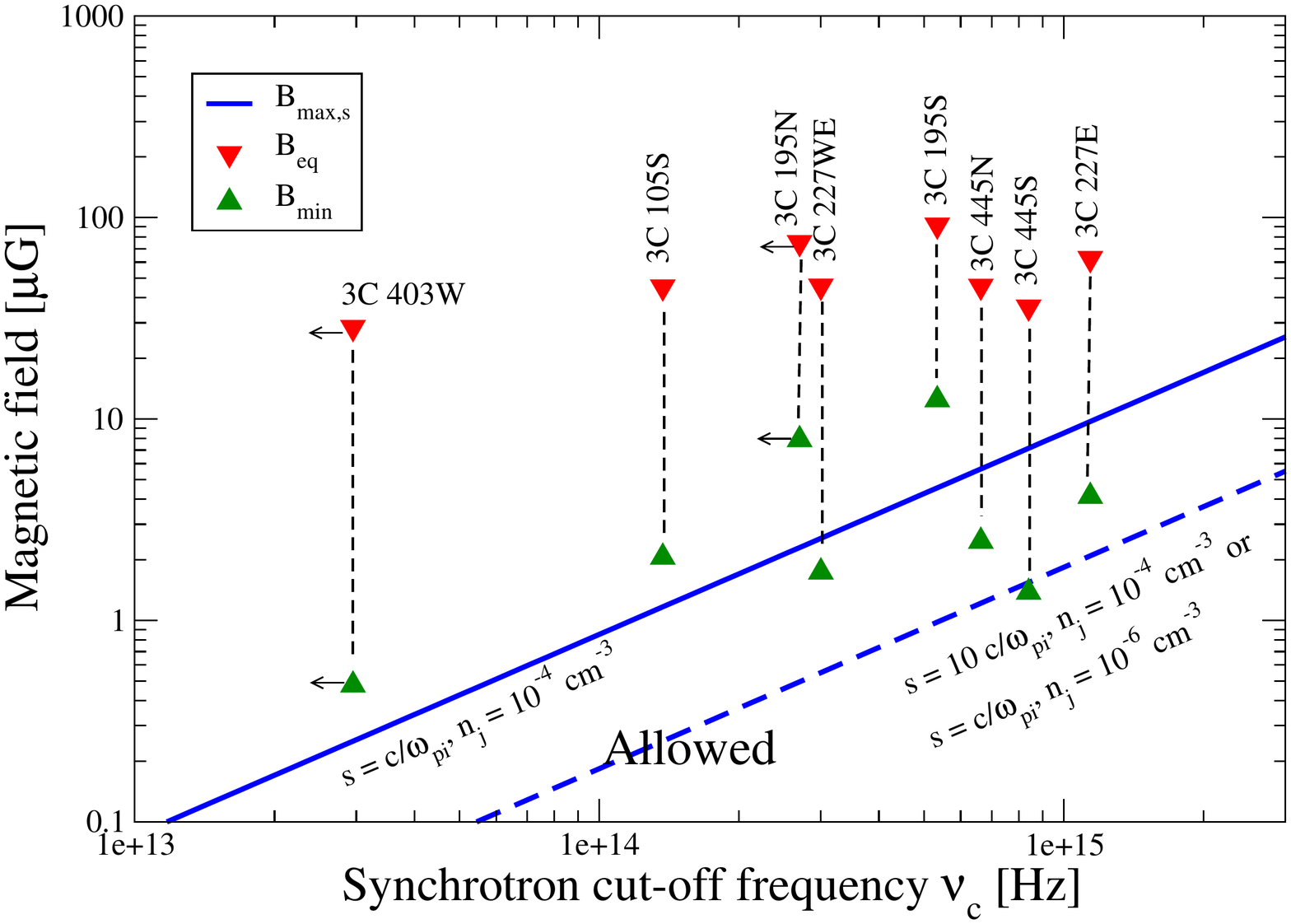}
\caption{Upper limit $B_{\rm max,s}$ for the magnetic field imposed by the 
condition  $\lambda_{\rm c,s} \ge \lambda_{\rm max}$ 
($n_{\rm j}=10^{-4}$~cm$^{-3}$: blue-solid line;
$n_{\rm j}=10^{-6}$~cm$^{-3}$: blue-dashed line). Triangles indicate the
maximum ($B_{\rm eq}$, red triangles down) and minimum ($B_{\rm min}$, 
green triangles up) field for the sources in Mack et al. (2009); see 
Table~\ref{tab_sources}.
\label{B_nu_sources}}
\end{figure}

If $\gamma_{\rm c}$ is determined by a competition between shock acceleration 
and synchrotron cooling (i.e. $t_{\rm acc} = t_{\rm synchr}$), 
the mean-free path of $\gamma_{\rm c}$-electrons is given by Eq.~(\ref{mfp_c}). 
By comparing  $\lambda_{\rm c,s}$ with the upper-limit $\lambda_{\rm max}$,
we find that  
\begin{eqnarray}
\begin{aligned}
\frac{\lambda_{\rm c,s}}{\lambda_{\rm max}} \sim & 3\times10^4
\left(\frac{v_{\rm sh}}{c/3}\right)^2
\left(\frac{\nu_{\rm c}}{10^{14}\,{\rm Hz}}\right)^{-\frac{3}{2}}
\left(\frac{B}{100\,\mu{\rm G}}\right)^{\frac{3}{2}}\\
&\left[\left(\frac{r}{7}\right)\left(\frac{n_{\rm j}}{10^{-4}\,{\rm cm^{-3}}}\right)\right]^{-\frac{1}{2}}.
\label{lambda_ratio}
\end{aligned}
\end{eqnarray}
Equivalently, setting $\lambda_{\rm c,s} \le \lambda_{\rm max}$ implies 
a magnetic field $B\le B_{\rm max,s}$, where
\begin{equation}\label{Bs}
\frac{B_{\rm max,s}}{\rm \mu G} \sim 0.8
\left(\frac{\nu_{\rm c}}{10^{14}\,{\rm Hz}}\right)
\left(\frac{v_{\rm sh}}{c/3}\right)^{-\frac{4}{3}}
\left[\left(\frac{r}{7}\right)\left(\frac{n_{\rm j}}{10^{-4}\,{\rm cm^{-3}}}\right)\right]^{\frac{1}{3}}.
\end{equation}
(Note that the same relationship is found by setting 
$\mathcal{D}_{\rm c,s} \le \mathcal{D}_{\rm max}$.)
In Fig.~\ref{B_nu_sources} we plot $B_{\rm max,s}$ for the cases of
$n_{\rm j} = 10^{-4}$ (blue-solid line) and $10^{-6}$~cm$^{-3}$ (blue-dashed line). 
The small values of $B_{\rm max,s}$ 
would require a very large energy density in non-thermal electrons in order to 
explain the synchrotron flux measured at radio-wavelengths. To demonstrate 
this, we consider the sample of hotspots observed at radio, IR and optical
frequencies by \cite{Mack_09}, and with a single radio-to-optical
spectral index $\alpha$, i.e. no spectral break (see Table~\ref{tab_sources}).

Non-thermal electrons follow a power-law energy distribution 
$N_e=K_e\gamma^{-p}$ with  $p = 2\alpha +1$ and minimum Lorentz 
factor\footnote{The value of minimum energy in non-thermal 
electrons $E_{\rm min} = \gamma_{\rm min}m_ec^2$  cannot be smaller than the energy 
 of the heated plasma downstream of the shock. By equating 
$n_{\rm j}m_pc^2/2 = 4n_{\rm j}K_{\rm B} T$, where $K_{\rm B}$ is the 
Boltzmann constant and $T$ is the temperature of the shocked jet,
we find that $K_{\rm B} T \sim m_pc^2/8\sim 0.1$~GeV and therefore 
$\gamma_{\rm min}$ has to be greater than 50.}
assumed to be $\gamma_{\rm min} = 100$. The electrons energy density is 
$U_e \sim K_e \gamma_{\rm min}^{2-p}/(p-2)$, where $K_e$
can be determined from the leptonic emission at a particular frequency. 
Considering the well resolved emission at $\nu=$8.4~GHz, with luminosity
$L_{8.4}$ emitted in a (cylinder-shaped) volume $V$
(see Table~\ref{tab_sources}), $U_e$ can be written as 
\begin{eqnarray}\label{Ue}
\begin{aligned}
\frac{U_{e}}{\rm erg\,cm^{-3}}  & \sim  10^{-9}
\left(\frac{p-2}{0.5}\right)^{-1}  
\left(\frac{\gamma_{\rm min}}{100}\right)^{2-p}
\left(\frac{\nu}{8.4\,{\rm GHz}}\right)^{\frac{p-3}{2}}\\
&\left(\frac{L_{8.4}}{10^{41}\,{\rm erg\,s^{-1}}}\right)
\left(\frac{V}{{\rm kpc^{3}}}\right)^{-1}
\left(\frac{B}{100\rm \mu G}\right)^{\frac{-p-1}{2}}.
\end{aligned}
\end{eqnarray}
The magnetic field is unknown, but we can set the upper-
and lower-limits. The former corresponds to the magnetic field in equipartition
with non-thermal particles. Setting $U_e(1+a)=B^2/(8\pi)$, where $a\ge 0$
takes into account the contribution of non-thermal protons,  we 
find that
\begin{eqnarray}\label{Beq}
\begin{aligned}
\frac{B_{\rm eq}}{\rm \mu G}  \sim &220^{\frac{7.5}{p+5}}
\left[(1+a)\left(\frac{p-2}{0.5}\right)^{-1} 
\left(\frac{\gamma_{\rm min}}{100}\right)^{2-p}\right.\\
&\left.\left(\frac{\nu}{8.4\,{\rm GHz}}\right)^{\frac{p-3}{2}}
\left(\frac{L_{8.4}}{10^{41}\,{\rm erg\,s^{-1}}}\right)
\left(\frac{V}{{\rm kpc^{3}}}\right)^{-1}\right]^{\frac{2}{p+5}}.
\end{aligned}
\end{eqnarray}
We calculate $B_{\rm eq}$ for all the sources in 
\cite{Mack_09} assuming $a=0$; see Table~\ref{tab_sources} and 
Fig.~\ref{B_nu_sources} 
(red-triangles down)\footnote{The equipartition field in Eq.~(\ref{Beq}) 
is slightly different from the value in \cite{Mack_09} given that we 
consider a cylinder-shaped volume, instead of an spheroid, and we set 
$a=0$ instead of 1. Note also that \cite{Mack_09} follow the approach of
\cite{Brunetti_97} to compute $B_{\rm eq}$.}.
Note that $B_{\rm eq}\sim 50$~$\mu$G in all the cases, and  far greater than 
$B_{\rm max,s}$ (blue-solid line), particularly  for those cases with 
$\nu_{\rm c} < 10^{15}$~Hz. 

\subsubsection{Minimum value of B}

In the extreme assumption that the non-thermal electron energy density is
$U_e = U_{\rm kin}$ (see Eq.~(\ref{U_kin})), the 
minimum value of the magnetic field required to emit a luminosity $L_{8.4}$ 
at frequency $\nu$ in a volume $V$ is
\begin{eqnarray}\label{Bmin}
\begin{aligned}
\frac{B_{\rm min}}{\rm \mu G}  & \sim  27^{\frac{3.5}{p+1}}
\left(\frac{\gamma_{\rm min}}{100}\right)^{\frac{4-2p}{(p+1)}}
\left(\frac{\nu}{8.4\,{\rm GHz}}\right)^{\frac{p-3}{p+1}}
\left(\frac{L_{8.4}}{10^{41}\,{\rm erg\,s^{-1}}}\right)^{\frac{2}{p+1}}\\
&\left[\left(\frac{\Gamma_{\rm j} - 1}{0.06}\right)
\left(\frac{p-2}{0.5}\right)
\left(\frac{V}{{\rm kpc^{3}}}\right)
\left(\frac{n_{\rm j}}{10^{-4}\rm cm^{-3}}\right)\right]^{\frac{-2}{p+1}}.
\end{aligned}
\end{eqnarray}
We compute $B_{\rm min,s}$ for all the sources in \cite{Mack_09}; see
Table~\ref{tab_sources} and Fig.~\ref{B_nu_sources} (green-triangles up).
We can see that $B_{\rm min} > B_{\rm max,s}$ (blue-solid line)
for those sources with $\nu_{\rm c} \lesssim 4\times10^{14}$~Hz 
(3C\,105S, 3C\,195N, 3C\,227WE
and 3C\,403W) whereas $B_{\rm min} < B_{\rm max,s}$
for hotspots with  $\nu_{\rm c} \gtrsim 4\times10^{14}$~Hz
(3C\,195S, 3C\,227E, 3C\,445N and 3C\,445S). Note however that:
\begin{itemize}
\item $n_{\rm j} \sim 10^{-4}$~cm$^{-3}$ is the upper limit found in Cygnus~A and
3C475, and 
therefore we expect values of  $B_{\rm min}$ greater than those 
plotted in Fig.~\ref{B_nu_sources} when the jet density is smaller than
$10^{-4}$~cm$^{-3}$ ($B_{\rm min} \propto n_{\rm j}^{-(p+1)/2}$). 
On the other hand, $B_{\rm max,s} \propto n_{\rm j}^{1/3}$ and therefore $B_{\rm max,s}$
decreases when smaller values of $n_{\rm j}$ are considered and
the ratio $B_{\rm min}/B_{\rm max,s} \propto n_{\rm j}^{-(p+5/6)}$. In particular,
the blue-dashed line in Fig~\ref{B_nu_sources} corresponds to the case of
$n_{\rm j} = 10^{-6}$~cm$^{-3}$ and $s = c/\omega_{\rm pi}$.   
In such a case,  sources 
3C\,195S, 3C\,227E, 3C\,445N and 3C\,445S move to the regime where
$B_{\rm min} > B_{\rm max,s}$. The minimum value of the jet density required to
match $B_{\rm min} = B_{\rm max,s}$ is 
listed in Table~\ref{tab_sources} for all the sources
considered in this paper. We can see for instance that the source 3C~195N 
necessitates $n_{\rm j} > 6.5\times10^{-4}$~cm$^{-3}$ to satisfy the 
condition $\lambda_{\rm c,s}<\lambda_{\rm max}$ and 
$U_e < U_{\rm kin}$.

\item Even when jets in FR galaxies are expected to be  
perpendicular to the line of sight, a small departure from the plane of the 
sky (i.e. $\theta_{\rm j}<90^{\circ}$) reduces the  size of the shock downstream 
region (see Eq.(\ref{L_theta})). In such a case, $B_{\rm min} \propto V^{-2/(p+1)}$ 
increases whereas $B_{\rm max,s}$ remains constant. 
Therefore, the situation is even more strongly ruled out when 
$\theta_{\rm j}<90^{\circ}$.
\end{itemize}

In the next section we show that even in the case that the extreme conditions
discussed before are assumed, the large value of the diffusion coefficient 
required for $\gamma_{\rm c}$ to be determined by synchrotron cooling 
cannot be explained in any well-established theoretical framework.

\begin{table*}
\caption{Physical parameters of the sources considered in this paper.
The redshift $z$, $\nu_{\rm c}$ and $\alpha$ are taken from Mack et al. (2009),
and $p=2\alpha+1$.
The synchrotron  specific luminosity at 8.4~GHz is calculated as 
$L_{8.4} = P_{8.4}\,10^7 \,8.4\times10^9$, where 
$P_{8.4}$  [W Hz$^{-1}$] is the measured power. 
The hotspot volume $V$  is calculated from the angular sizes 
tabulated in Table~4 of Mack et al. (2009) together with 
$P_{8.4}$.}
\centering
\label{tab_sources}
\begin{tabular}{lcrc|ccr|rrrc}
\hline\hline
Source&$z$&$\nu_{\rm c}$&$\alpha$& $p$& $L_{8.4}$&$V$ &$B_{\rm eq}$& $B_{\rm min}$ &$B_{\rm max,s}$ & $n_{\rm j,min}$\\
& &[$10^{14}$~Hz]&&&[erg/s]&[kpc$^3$] &[$\mu$G]&[$\mu$G]&[$\mu$G] &[cm$^{-3}$]\\
\hline
3C\,105S &0.089& 1.37  &0.75 &2.5 &1.42$\times$10$^{42}$& 1205.63& 45.27& 
2.06&1.16&1.92$\times$10$^{-4}$\\ 
3C\,195N &0.110& $<$2.70&0.95 &2.9 &1.15$\times$10$^{41}$& 38.12& 75.11& 7.89&
2.30&6.51$\times$10$^{-4}$\\
3C\,195S &0.110 & 5.34  &1.00 &3.0 &1.71$\times$10$^{41}$& 33.58& 91.76&  12.45&
4.55&3.42$\times$10$^{-4}$\\
3C\,227WE&0.086& 3.00  &0.65 &2.3 &3.19$\times$10$^{40}$& 19.26& 45.63&  1.74&
2.55&6.78$\times$10$^{-5}$\\
3C\,227E &0.086& 11.4  &0.75 &2.5 &7.14$\times$10$^{40}$& 17.99& 62.60 &  4.12&
9.71&3.96$\times$10$^{-5}$\\
3C\,403W &0.059& $<$0.29&0.55&2.1 &3.95$\times$10$^{40}$& 167.9& 28.46 &  0.48&
0.25&1.96$\times$10$^{-4}$\\
3C\,445N &0.056& 6.63  &0.85 &2.7 &2.18$\times$10$^{40}$& 29.36& 45.60&  2.47&
5.65&3.97$\times$10$^{-5}$\\
3C\,445S &0.056& 8.40  &0.80 &2.6 &5.04$\times$10$^{40}$& 139.42& 35.94 &  1.38&
7.15&1.60$\times$10$^{-5}$\\
\hline\hline
\end{tabular}
\end{table*}

\section{Particle acceleration and magnetic field amplification}
\label{plasma}

The diffusion coefficient resulting from the assumption that
$\gamma_{\rm c}$  is determined by synchrotron cooling is 
very large, $\mathcal{D}_{\rm c,s}/\mathcal{D}_{\rm Bohm}\sim 10^6$-$10^7$, as we
show in Eq.~(\ref{DDBohm}).
For comparison, this is $\sim$10$^3$-10$^4$ times 
larger than $\mathcal{D}/\mathcal{D_{\rm Bohm}}$ for TeV 
particles diffusing through the Galactic interstellar medium.
It is even more extreme when 
compared with  $\mathcal{D} \sim \mathcal{D}_{\rm Bohm}$ during diffusive 
shock acceleration in supernova remnants 
where the magnetic field is strongly amplified by the non-resonant hybrid (NRH) 
instability \citep{Bell_04} and structured on the scale 
of the cosmic ray (CR) Larmor radius.

In non-relativistic shocks, the condition for the NRH instability to be 
active is that the upstream magnetic energy density must be less than 
$\eta U_{\rm kin} (v_{\rm sh}/c)$,
where $\eta$ is the efficiency with which the available kinetic energy 
is given to CR (see Sect.~\ref{hotspots_mfd} and Appendix~\ref{ap}).  
This condition is easily met in hotspots but it may 
not apply to relativistic shocks.
One possible difference is that magnetic field amplification at relativistic 
shocks might be driven only by mildly relativistic particles since 
CR spectra at relativistic shocks are relatively steep with the CR energy 
density dominated by low energy CR. Fully developed
magnetic turbulence on the scale of the GeV Larmor radius
would naturally scatter TeV particles with 
$\mathcal{D}/\mathcal{D_{\rm Bohm}} \sim 10^3$ since 
$\mathcal{D}/\mathcal{D_{\rm Bohm}} \sim r_{\rm g}/s$, as we will see in 
Sect.~\ref{hotspots_mfd}.
However, $\mathcal{D}/\mathcal{D_{\rm Bohm}} \sim 10^3$ is not sufficient to 
explain  spectral turnover in the range $10^{14}-10^{15}$~Hz (implying
$\mathcal{D}_{\rm c,s}/\mathcal{D}_{\rm Bohm}\sim 10^6-10^7$)
and the NRH instability must be 
ruled out if we assume that the turnover is due to synchrotron losses.

In ultra-relativistic shocks in weakly magnetised plasmas
($\sigma_{\rm j} < 10^{-3}$), the Weibel instability 
dominates and generates magnetic field on the small scale of the ion 
collisionless skin depth $c/\omega_{\rm pi}$. \cite{Sironi_11} found that
the amplified magnetic field has a scalelength of 
$\sim 10 c/\omega_{pi}$ but the factor 10 may be due to their shock Lorentz 
factor $\Gamma_{\rm j}=15$ which increases $c/\omega_{\rm pi}$ by 
$\sim$$\sqrt{\Gamma_{\rm j}}$ when the relativistic ion mass is allowed for
(see Eq.~(\ref{c_omega_pi})).
If we assume fully developed Weibel turbulence with CR scattered by randomly 
orientated magnetic cells on a scale $c/\omega_{\rm pi}$, 
the diffusion coefficient is given by 
$\mathcal{D}_{\rm max}/\mathcal{D}_{\rm Bohm} = r_{\rm g}(\gamma_{\rm c})/(c/\omega_{\rm pi}) \sim 3\times10^4$ as shown in Eq.~(\ref{D_DBohm}).
This value of the diffusion coefficient is large but still much smaller 
than $\mathcal{D}_{\rm c,s}/\mathcal{D}_{\rm Bohm}\sim 10^6$-$10^7$ 
that would be required to explain the spectral turnover at 
$\nu_{\rm c}=10^{14}$~Hz in a jet with density $n_{\rm j}=10^{-4}$~cm$^{-3}$, 
at least  $B \le B_{\rm max,s}$.
A further difficulty with a Weibel scenario is that post-shock Weibel 
turbulence decays on a scale of 
$\sim 10^3 c/\omega_{\rm pi} \sim 10^{-5}\sqrt{\Gamma_{\rm j}}(n_{\rm j}/10^{-4}{\rm cm^{-3}})^{-0.5}$~pc 
\citep{Sironi_11} which is many orders of magnitude 
smaller than the size of the hotspot, which is of the order of 
10~pc to kpc. We note that the same discrepancy is found in gamma-ray bursts 
\cite[e.g][]{Gruzinov_99,Peer_06}, although it is not 
completely clear at present how the
small-scale magnetic turbulence evolves downstream of the shock 
\cite[see e.g.][]{Sironi_rev}.

Fully developed turbulence with a magnetic field of $\sim$100~$\mu$G cannot be 
responsible for $\mathcal{D}_{\rm c,s}/\mathcal{D}_{\rm Bohm} \sim 10^6$-$10^7$ 
even if its cell size were as small as $c/\omega_{\rm pi}$, as is shown in 
Fig.~\ref{B_nu_sources} and Eq.~(\ref{D_DBohm}).  There remains the possibility 
that the magnetic field might consist of a long scalelength component 
with $B \sim$100~$\mu$G with a small-scale perturbation 
$\delta B \ll B$. Magnetic turbulence $\delta B \ll B$ 
probably occurs in the Galaxy where CR drift along 
relatively well-ordered magnetic field lines with weak scattering by 
Alfven waves with amplitude $\delta B$ driven by CR drifts at the order of the 
Alfven speed.
Under these conditions the CR current is too weak to drive the NRH instability.
The Alfven waves are driven resonantly with a wavelength similar to the CR 
Larmor radius, and $\mathcal{D}/\mathcal{D_{\rm Bohm}} \sim (\delta B/B)^{-2}$ 
requires fluctuations in the  magnetic field as small as
$\delta B/B \sim 10^{-3}$ in order to reach $\mathcal{D}_{\rm c,s}/\mathcal{D_{\rm Bohm}}\sim 10^6$-$10^7$.
Note however that even in the very weak CR drifts in the interstellar medium 
$\mathcal{D}/\mathcal{D_{\rm Bohm}} \sim 10^3$.
For this scenario to hold for hotspots, a valid theory would need to explain 
how magnetic field $B$ could be amplified to
$\sim$100~$\mu$G on scales larger than a CR Larmor radius while producing 
fluctuations $\delta B$ on a Larmor 
scale with only $\delta B/B \sim 10^{-3}$ at a relative amplitude smaller
than that found in the Galactic interstellar medium.

Although it is impossible to rule out all possibilities, 
it appears extremely difficult to construct a scenario in which 
$\mathcal{D}_{\rm c,s}/\mathcal{D_{\rm Bohm}}\sim 10^6$-$10^7$,
as required by the supposition that the IR/optical turnover
in the synchrotron spectrum is caused by synchrotron radiation losses.
We therefore suggest that the cut-off in the spectrum has a different 
cause, which we now explore in the next section.

\section{Non-resonant hybrid instabilities in mildly relativistic shocks}
\label{hotspots_mfd}

If radiative (synchrotron) losses are not relevant to determining the
maximum energy, then this maximum energy must ultimately
determined by the ability to scatter particles downstream of the shock.
We explore the possibility that the maximum energy achieved
by electrons in the jet reverse shock is constrained by 
magnetic turbulence generated by low energy CRs in perpendicular shocks.

We consider that the amplified hotspot magnetic field $B$ is turbulent, 
and that the large-scale background field downstream of the reverse shock 
is $B_{\rm jd}$ nearly perpendicular to the 
shock normal because the perpendicular component is compressed
and enhanced by a factor of $4$ to $7$ (i.e. $B_{\rm jd}\sim r B_{\rm j}$). 
In such a case, to accelerate particles  up to an energy 
$E_{\rm c}$ via a diffusive mechanism,  the mean-free path 
$\lambda_{\rm c}\sim r_{\rm g}(\gamma_{\rm c},B)^2/s$
in the shock downstream region, where $B$ is a small-scale field, 
has to be smaller than Larmor radius in $B_{\rm jd}$ 
\citep{lemoine-pelletier-10,Brian_14}\footnote{When the mean-free path 
of particles in the turbulent field
exceeds the Larmor radius in the background compressed field, particles
return to helical orbits and diffusion ceases.}. 
The condition $\lambda_{\rm c} \lesssim r_{\rm g}(\gamma_{\rm c},B_{\rm jd})$
is satisfied when the magnetic-turbulence scale-length is
\begin{equation}\label{s_nrh}
s \ge \frac{E_{\rm c}}{eB}\left(\frac{B_{\rm jd}}{B}\right) =  
r_{\rm g}(\gamma_{\rm s},B),
\end{equation}
where $r_{\rm g}(\gamma_{\rm s},B)$ is the Larmor radius of protons with energy
\begin{eqnarray}\label{gamma_nrh}
\begin{aligned}
E_{\rm s} &= E_{\rm c}\left(\frac{B_{\rm jd}}{B}\right) =
0.07 E_{\rm c}\left(\frac{r}{7}\right)
\left(\frac{B_{\rm j}}{\rm \mu G}\right)
\left(\frac{B}{100 \rm \mu G}\right)^{-1}\\
&\sim10 \left(\frac{r}{7}\right)
\left(\frac{\nu_{\rm c}}{10^{14}\,{\rm Hz}}\right)^{\frac{1}{2}}
\left(\frac{B_{\rm jd}}{\rm \mu G}\right)
\left(\frac{B}{\rm 100\mu G}\right)^{-\frac{5}{2}}\,\,{\rm GeV},
\end{aligned}
\end{eqnarray}
where we take $B\sim100$~$\mu$G and $B_{\rm j}\sim \mu$G as characteristic 
values. Note that  
\begin{equation}\label{s}
\frac{s}{\rm cm} > 5\times10^{11}\left(\frac{r}{7}\right)
\left(\frac{\nu_{\rm c}}{10^{14}\,{\rm Hz}}\right)^{\frac{1}{2}}
\left(\frac{B_{\rm jd}}{\rm \mu G}\right)
\left(\frac{B}{\rm 100\mu G}\right)^{-\frac{5}{2}}
\end{equation}
is greater than $c/\omega_{\rm pi}$ in Eq.~(\ref{c_omega_pi}), as 
required. Note however that this limit, $s\gtrsim 500 \,c/\omega_{\rm pi}$
for typical values considered in this paper, cannot be fulfilled by 
Weibel-generated turbulence with scale $\sim c/\omega_{\rm pi}$. Therefore,
the maximum energy achieved by electrons in the jet reverse
shock, $E_{\rm c}$, cannot be constrained by  Weibel instabilities.

Turbulence on a scale greater than $c/\omega_{\rm pi}$ may be excited through
the non-resonant  hybrid (NRH) instability by the diamagnetic drift 
of CR on either side of the shock. 
In the simplest form of the NRH instability \citep{Bell_04,Tony_05}, 
the CR Larmor radius 
in the unperturbed background field is much greater than the 
wavelength of field perturbations  and therefore the streaming
of CRs carrying the electric current $j_{\rm cr}$
is undeflected. The force -$\vec j_{\rm cr}\times \vec B$
acts to expand loops in the magnetic field, and therefore $B$ 
increases. This produces an increment in 
-$\vec j_{\rm cr}\times \vec B$ and generates a positive feedback loop 
that drives the NRH instability and amplifies the magnetic field.
For the diamagnetic drift in the plane of the shock to amplify
the magnetic field (see Appendix~\ref{ap})
the NRH growth rate has to be sufficient for the instability 
to grow through $\sim$10 e-foldings at the maximum growth rate 
$\Gamma_{\rm max}$ \citep{Bell_04,Bell_rev_14} in the time the plasma 
flows through a distance $r_{\rm g}(B_{\rm js})$ in the downstream region,
where $r_{\rm g}(B_{\rm js})$ 
is the Larmor radius in the ordered field $B_{\rm js}$. That is, the condition
$\Gamma_{\rm max}\,r_{\rm g}(B_{\rm js})/v_{\rm d} > 10$ must be satisfied
(see Appendix~\ref{ap}). If the field is strongly amplified, the instability 
can be expected
to saturate when its characteristic scale grows to the Larmor radius of the
CR driving the instability. Thus, $s$ in Eq.~(\ref{s_nrh}) can be expected to
match the Larmor radius of the highest energy CR driving the instability.
If these CR have an energy $E_{\rm nrh}$, then  $E_{\rm nrh}\sim E_{\rm s}$.
From Eq.~(\ref{gamma_nrh}), if $v_{\rm sh} \sim c/3$ then CR with energy 
$E_{\rm nrh}$ correspond
to mildly supra-thermal protons ($E_{\rm nrh}\sim 100\,m_pv_{\rm sh}^2$)
in the downstream plasma. It is entirely reasonable that protons with this
energy should be present in large numbers downstream of the shock and
drive the NRH instability.

In order to check that there is enough energy in $E_{\rm nrh}$-protons
to excite the non-resonant turbulence, we consider whether the number of 
e-foldings required to amplify the magnetic field up to the saturation value 
is of the order of 10 \citep{Bell_04, Bell_rev_14}. The condition 
for efficient magnetic field amplification by NRH instabilities is that
$\Gamma_{\rm max}\,r_{\rm g}(B_{\rm js})/v_{\rm d} > 10$, as explained above.
This condition leads to 
\begin{eqnarray}
\begin{aligned}
\eta &> 10 \,r^{3/2}\,\sqrt{\sigma_{\rm j}} \\
&\sim 0.04 \left(\frac{r}{7}\right)^{\frac{3}{2}}\left(\frac{B_{\rm j}}{\mu\rm G}\right)
\left(\frac{\Gamma_{\rm j} - 1}{0.06}\right)^{-\frac{1}{2}} 
\left(\frac{n_{\rm j}}{10^{-4}\,{\rm cm^{-3}}}\right)^{-\frac{1}{2}}
\end{aligned}
\end{eqnarray}
(see Appendix~\ref{ap}), where $\eta\propto P_{\rm CR}$ is the acceleration efficiency
and $\sigma_{\rm j}$ is the jet magnetisation parameter defined in 
Eq.~(\ref{sigma}). Given that 
particles accelerated in relativistic shocks follow a power-law energy
distribution  steeper than the canonical distribution, 
the CR pressure $P_{\rm CR}$ is dominated by low energy particles.  
Therefore, the condition for NRH instability growth is that the acceleration 
efficiency of low energy CR has to be
$\eta \sim 0.04$ for characteristic values considered in this paper. 
Such a value of $\eta$ is very reasonable. For comparison, CR acceleration 
in supernova remnants is usually thought to be in the range 10\%-50\%.

From these  estimations we can conclude that NRH instabilities 
generated by CRs with energies $\lesssim E_{\rm nrh}$ can grow fast enough 
to amplify the jet magnetic field 
from $\sim$1 to 100~$\mu$G and accelerate particles up to energies
$\sim$$E_{\rm c}$ observed in the hotspots of FR~II radiogalaxies.
The advantage of magnetic turbulence being generated by CR current is that
the amplified magnetic field persists over long distances downstream of the 
shock, and therefore particles accelerated very near the shock can emit
synchrotron radiation far downstream. 
This framework also applies to hotspots with break in the synchrotron 
spectrum, and we will explore this situation in depth in a following paper. 
 
\section{Summary and conclusions}
\label{disc}

Motivated by the recent realisation of magnetic field damping in the 
southern hotspot of the radiogalaxy 4C74.26 \citep{4c7426}, we have explored 
in great depth the physical conditions in the hotspots of 
a larger number of FR~II radiogalaxies. 
In particular, we have investigated the physical mechanism that constraints the
maximum energy of particles accelerated at the jet reverse shock. 

Based on one observable (the cut-off $\nu_{\rm c}$ of the synchrotron spectrum)
and one physical requirement ($s \ge c/\omega_{\rm pi}$)
we have found that extreme conditions in the jet plasma would be required for 
$\nu_{\rm c}\sim 10^{14}$-$10^{15}$~Hz to be determined by synchrotron
cooling, as usually assumed.
By equating the acceleration and synchrotron cooling timescales,  
the mean free path of $\nu_{\rm c}$-synchrotron emitting electrons 
is greater than the maximum value $r_{\rm g}^2/(c/\omega_{\rm pi})$
imposed by plasma physics for reasonable values of the magnetic field and
jet density (see Eq.~(\ref{lambda_ratio})).
By  considering a sample of 8 hotspots observed with high spatial
resolution at optical, IR and radio wavelengths \citep{Mack_09}, we show that 
unreasonably large values of the jet density would be required 
(see Table~\ref{tab_sources}) to explain the synchrotron flux at 8.4~GHz
when $E_{\rm c}$ (maximum energy of non-thermal electrons) is determined 
by synchrotron cooling (see Fig.~\ref{B_nu_sources}). The key  steps in
our argument are outlined in Table~\ref{argument}. 

As mentioned in Sect.~\ref{plasma}, the
structure of the magnetic field downstream of the shock is not completely
understood at the moment. Weibel-mediated shocks generate the magnetic field 
and accelerate particles \cite[e.g.][]{Spitkovsky_08II,Martins_09}. 
However, the characteristic scale of Weibel turbulence
cannot account for the cut-off of the synchrotron spectrum observed in 
hotspots because this scale size is too small., nor  the large extent 
of the hotspot synchrotron emission, 
much larger than the magnetic decay of $\sim 100c/\omega_{\rm pi}$ predicted
by numerical calculations. A viable alternative is that
turbulence is generated by the streaming of CRs with energy 
$E_{\rm nrh}\sim E_{\rm c}B_{\rm jd}/B \sim 0.01 E_{\rm c}$ 
(see Sect.~\ref{hotspots_mfd}). The amplified magnetic 
field has a scale-length of the order of the Larmor radius of 
$E_{\rm nrh}$-protons and persists over long distances downstream of the 
shock, accounting for the extent of the synchrotron emitting hotspot. 

In a future work, we will apply our arguments
to the  very well known sources Cygnus~A and 3C445 
for which well resolved and multi-wavelength data are  
available \cite[e.g.][]{sources,Pyrzas_15}. 
By modelling the particle acceleration and transport 
downstream the shock 
we will be able to determine the details of the
magnetic field structure downstream of mildly relativistic shocks.

\section*{Acknowledgements}

The authors would like to thank the referee for a constructive report, and 
M.A. Prieto for providing useful 
information about the sources considered in this paper. 
A.T.A. thanks M. Perucho, W. Potter and L. Sironi for useful discussions 
about jet physics. 
The research leading to this article has received funding
from the European Research Council under the European
Community's Seventh Framework Programme (FP7/2007-2013)/ERC grant agreement 
no. 247039.
We acknowledge support from the UK
Science and Technology Facilities Council under grant No. ST/K00106X/1.

\bibliographystyle{mnras}
\bibliography{biblio_CRs} 

\appendix

\section{Condition for efficient NRH instability in hotspots}
\label{ap}

As noted in Sect.~\ref{hotspots_mfd}, a condition for effective CR scattering 
by turbulent magnetic fields amplified by the NRH instability with
maximum growth rate $\Gamma_{\rm max}$ is that 
$\Gamma_{\rm max}r_{\rm g}/v_{\rm d} > 10$ where $v_{\rm d}$ is the downstream 
flow velocity 
and $r_g$ is the Larmor radius of the CR driving the instability.  
The perpendicular component of the magnetic field in the jet is compressed 
by the shock producing a downstream field that is predominantly perpendicular 
on the large scale.  Drift of CR along the shock surface produces a 
diamagnetic current  that can drive the NRH instability.   
The CR current $j_{\rm CR}$ is perpendicular to both the shock normal and the 
large scale 
magnetic field and  extends a distance $\sim r_{\rm g}$ downstream of the shock. 
The NRH instability must be driven through many e-foldings during the 
time $t_{\rm amp}\sim r_{\rm g}/v_{\rm d}$ during which a fluid element is
subject to the diamagnetic current.
In this configuration the NRH growth rate is smaller by a numerical factor
of order one than in the case of aligned currents and magnetic field
\citep{Tony_05}. However,
$\Gamma_{\rm max} \sim j_{\rm CR} \sqrt{4\pi/\rho_{\rm jd}}$
is still a good order-of-magnitude measure of the growth rate,
where $\rho_{\rm jd} \sim r\, m_pn_{\rm j}$ is the 
density in the shock downstream region. The condition
$\Gamma_{\rm max} t_{\rm amp} \sim 10$ provides a good estimate of the time
$t_{\rm amp}$ for strong non-linear amplification, giving
\begin{equation}
\sqrt{\frac{4\pi}{\rho_{\rm jd} v_{\rm d}^2}}  \int j_{\rm CR}\, {\rm d}z > 10,
\end{equation}
where $j_{\rm CR}$ depends on distance $z$ from the shock.
From the momentum equation the downstream CR pressure $P_{CR}$ must be 
balanced by the magnetic force: 
$\int j_{\rm CR} B_{\rm jd} \,{\rm d}z \approx P_{\rm CR}$
giving the condition 
\begin{equation}
\sqrt{\frac{4\pi}{\rho_{\rm jd} v_{\rm d}^2}} \left(\frac{P_{\rm CR}}{B_{\rm jd}}\right) 
> 10,
\end{equation}
or equivalently
\begin{equation}
\label{a3}
\eta = \frac{P_{CR}}{\rho_{\rm jd} v_{\rm d}^2} > 10
\left (\frac { B_{\rm jd}^2/4\pi}{\rho_{\rm jd} v_{\rm d}^2} \right )^{1/2} = 
10 \, r^{3/2}\,\sqrt{\sigma_{\rm j}}
\end{equation}
where $\eta = P_{\rm CR}/{\rho v_{\rm d}^2}$ is the CR acceleration efficiency, 
as quoted in Sect.~\ref{hotspots_mfd}. Equation~\ref{a3} is thus the 
condition for efficient NRH instability in jet reverse shocks.

\bsp	
\label{lastpage}
\end{document}